\documentclass[twocolumn,showpacs,preprintnumbers,amsmath,amssymb]{revtex4}

\usepackage{graphicx}
\usepackage{dcolumn}
\usepackage{bm}

\begin{document}

\title{A Dynamic Model of Streamer Coupling for High Pressure Discharges}
\author{Qing Li$^1$}
\author{Demetre J. Economou$^2$} \author{Yi-Kang Pu$^1$}
 \affiliation{%
$^1$Department of Engineering Physics, Tsinghua University, Beijing
100084, China
}%
\affiliation{
$^2$Department of Chemical and Biomolecular
Engineering, Plasma Processing Laboratory, University of Houston,
Houston, Texas 77204-4004}


\begin{abstract}
A streamer coupling theory is developed to describe the formation of
homogenous emission, and the high moving speed of emission patterns
in high pressure discharges. By considering the effects of both
electron diffusion and electronic drift in the streamer head, the
minimum required preionization level $n_{\rm min}$ for the formation
of streamer coupling is found to depend on electric field strength,
gas pressure and electron temperature. The homogeneity and moving
speed of the emission pattern in streamer coupling head increase
with preionization level $n_0$, when $n_0 > n_{\rm min}$. The
predicted results for atmospheric helium plasma indicate $n_{\rm
min} \sim 10^5~{\rm cm^{-3}}$ and moving speed of $10^4 - 10^6$ m/s,
in agreement with experiments.
\end{abstract}

\pacs{51.50.+v, 52.80.Dy, 52.80.Hc, 52.80.Mg}
\maketitle

Homogeneous or glow-like emission in high pressure discharges has
been studied for decades, and most of the works focused on
experimental conditions for their formation and methods of improving
discharge homogeneity \cite{glow98}. Although high preionization
level is verified to be necessary for homogeneous discharges
\cite{glow98,DBD02,glow09}, its formation mechanism, different from
those of glow discharges in low pressure and streamer discharges in
high pressure, has not been fully clarified \cite{glow09,Raizer91}.
In a recent experiment, atmospheric pressure plasma jet (APPJ),
originating from dielectric barrier discharge (DBD) and spraying
into ambient air, is found to be a traveling bulletlike plasma
volume with a high moving speed of the order of $10^4 \sim 10^6$ m/s
and a ring-shaped cross-sectional emission pattern, named as "plasma
bullet" \cite{bullet05,Lu06,bullet10,La09,David10}. A model of
self-sustained photo-ionization streamer \cite{Lu06}, first
developed by Dawson and Winn \cite{streamer65}, was invoked to
explain the nature of plasma bullet as one single streamer
\cite{streamer65,streamer96}. Although the model sufficed to account
for the scale of moving speed, it was inadequate to explain the
ring-shaped emission pattern of axial-symmetrical homogeneity and
the change of moving speed with external electric field
\cite{bullet10,La09,David10}. Thus, we suppose that APPJ is neither
a single streamer discharge, nor traditional Townsend or glow
discharges in low pressure, which are homogeneous in radial
direction and brightest in the vicinity of the anode or cathode
\cite{glow98,DBD02,glow09}.The discharge mechanism should be a new
one, which we here call it streamer coupling.

The thoughts of streamer coupling model was primarily advocated by
Palmer to predict a volume-stabilized glow-like discharge in
atmospheric pressure helium discharge \cite{coupling74}. In Palmer's
theory, the interaction of simultaneous developing streamers leads
to the formation of one large discharge canal, and the dominating
force in each streamer head is electron diffusion
\cite{coupling01,coupling80}. However, experimental results show
that, in high pressure experiments such as atmospheric pressure
discharge, the dominating force responsible for the electron cloud
expansion in a streamer head is the electrostatic repulsion of
high-density charged particles, instead of the diffusion caused by
electron density gradient \cite{Raizer98}. Therefore in Palmer's
model the predicted minimum preionization level of $10^4~{\rm
cm^{-3}}$ for homogeneous discharge \cite{coupling74} is not
consistent with the experimental value of $10^5~{\rm cm^{-3}}$
\cite{coupling80}. Another defect in Palmer's model is that there is
no explicit relationship between important physical properties of
the discharge gas with the formation of homogeneous discharge, such
as electric field and gas pressure. In this work, by considering the
electron diffusion, electrostatic repulsion in streamer head, and
electron drift under the electric field, an improved streamer
coupling model is proposed to describe the dynamics of high pressure
discharge patterns.



We consider the "fluid approximation" for each streamer.
For the sake of simplicity, we here investigate only the primary
anode-directed streamer and assume it propagates in a uniform
background electric field $E_0$. The continuity equation for the
basic dynamics of a streamer formation and propagation is,
\begin{equation}
\frac{\partial{n_e}}{\partial{t}} = -
\nabla\cdot({n_e}\vec{\upsilon}_e) + S_e ,
\end{equation}
where $n_e$ is the electron density, $S_e$ is the electron source
term, and $\vec{\upsilon}_e$ is the electron velocity, determined by
\begin{equation}\label{eq:speed}
\vec{\upsilon}_e = - \mu_e\vec{E} - \frac{D_e}{n_e}\nabla{n_e},
\end{equation}
with $\mu_e$ and $D_e$ are the electron mobility and the diffusion
coefficient, respectively. A detailed analysis responsible for these
equations is given as follows.

(1) Resulting from drift and diffusion of charged particles in the
local electric field $\vec{E}$, streamer propagation is mainly
determined by the motion of electrons. The ions can be treated as
immovable particles, since their mobility $\mu_i$ and diffusion
coefficient $D_i$ can actually be neglected, comparing with those of
electrons \cite{Raizer98, streamer94, streamer02}.

(2) The source term $S_e$ can be treated in two processes. In the
process of primary avalanche, the source term $S_e$ is directly
proportional to the exponent of first Townsend ionization rate
$\alpha$, as $S_e \propto e^{\alpha{x}}$, where $x$ is the length of
the avalanche \cite{Raizer98}. In the other process of streamer
propagation, the source term $S_e$ could be treated as $S_e \propto
Ee^{-|E_0/E|}$, where $E_0$ and $E$ are external and total electric
fields, respectively \cite{streamer96}. In the present work, we
focus on the former case.

(3) We concentrate on the streamer dynamics under the strong
external electric field $\vec{E}_0$, as in high pressure discharges.
The criterion of streamer formation says, a streamer is born of an
avalanche if the electric field $E^\prime$ induced from the space
charge in the streamer head reaches the order of external field
$E_0$ \cite{Raizer98}. The correspondingly approximate equality is,
\begin{equation}\label{eq:Ecriterion}
E^\prime = 2\frac{e}{4\pi\varepsilon_0R_0^2}e^{\alpha{x_0}} = E_0,
\end{equation}
where $R_0$ is the characteristic radius of space charge in the
streamer head at the transformation point. The streamer head region
of intensive ionization, moving together with a strong field $E =
E_0 + E^\prime$, transforms the gas to plasma. A plasma channel is
left due to the production of new plasma region.

\begin{figure}
\includegraphics[angle=0, width=0.25\textwidth]{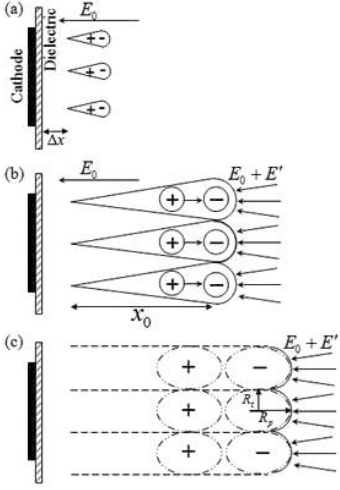}
\caption{\label{fig:streamer} Simultaneous avalanche-streamer
transition. (a) Avalanches start to develop when electric field
arises to a certain value $E_0$. (b) Primary avalanches turn into
streamers when they satisfy the criterion of streamer formation at
the length of $x_0$. (c) Adjacent streamers just overlap with each
other to form streamer coupling. }
\end{figure}

For the case of high pressure discharges in capacitively coupled
plasmas, simultaneous electrons leave the dielectric surface coated
on the instantaneous cathode towards the anode \cite{DBD02,glow09},
when the electrode polarity connected to the squared power source
turns from positive to negative half-cycle at time $t = -\Delta t$.
$\Delta t$ is assumed to be the arising time of external electric
field from 0 to the value at which the primary avalanche starts at
the leaving place of $x = \Delta{x}$, the distance to dielectric
surface. After the time $t = 0$, the supplying pulsed power source
is assumed to be sustained at constant value $E_0$ for the streamer
propagation [see Fig.~\ref{fig:streamer}(a)]. At a certain place $x
= x_0$ with $t = t_0$, these simultaneous primary avalanches
transform into simultaneous streamers. The streamer head radii in
propagating and transverse directions are assumed to be $R_p$ and
$R_t$, respectively [see Figs.~\ref{fig:streamer}(b) and (c)]. These
simultaneous streamers will overlap if the transverse radius $R_t$
is larger than the half distance between adjacent streamer head
centers $(4\pi{n_0}/3)^{-1/3}$, i.e.,
\begin{equation}\label{eq:nr}
n_0 \geq \frac{3}{4\pi{R_t}^3}.
\end{equation}
For better discussion bellow, the surface density of simultaneous
primary electrons is replaced by volume density $n_0$, which is
known as preionization level. The equal sign case of
Eq.~(\ref{eq:nr}) is shown in Fig.~\ref{fig:streamer}(c). The
overlapping streamers hereafter is called streamer coupling, whose
propagation likes one single streamer, except that it has a much
larger space charge head.

For the avalanche-streamer transition, if the expansive force at
streamer head edge is dominated by electrostatic repulsion or
electron diffusion, according to Eq.~(\ref{eq:speed}), the speed of
electron can be expressed as:
\begin{equation}\label{eq:largespeed}
{\upsilon}_e \approx \max\Big[\mu_e|\vec{E}_0 + \vec{E}^\prime|,
\Big|-\mu_e\vec{E}_0-\frac{D_e}{n_e}\nabla{n_e}\Big| \Big].
\end{equation}
In experimental breakdown condition of high pressure discharges, the
expansion of an avalanche head is mainly due to repulsive force
rather than diffusion one. The difference of the two forces in
magnitude can be one or two orders in many cases, such as in
atmospheric pressure air discharge \cite{Raizer98}.

Due to the cancelation of induced repulsive field between the
simultaneously developed adjacent streamer heads in transverse
direction, the dominator of Eq.~(\ref{eq:largespeed}) for the
directions of propagation and transverse are repulsion and
diffusion, respectively. Using Eq.~(\ref{eq:Ecriterion}) to the
propagating direction along the avalanche development, we obtain:
\begin{equation}\label{eq:Rp}
R_p \approx \frac{3}{2\alpha(E_0)}.
\end{equation}
For the sake of simplicity, the assumption $\alpha = \alpha(E_0)$ is
used in Eq.~(\ref{eq:Rp}) during the development of an avalanche
when the external field is only sightly distorted \cite{Raizer98}.
Also the corresponding transverse radius $R_t$ increased by
diffusion is:
\begin{equation}\label{eq:Rt}
R_t \approx
\Big[\frac{4D_e}{\mu_eE_0\alpha(E_0)}\ln\frac{9\pi\varepsilon_0E_0}{2e\alpha^2(E_0)}\Big]^{1/2},
\end{equation}
where $E \approx E_0$ is used from $x = 0$ to $x = x_0$ in
Eqs.~(\ref{eq:Rp}) and~(\ref{eq:Rt}) in the avalanche development. A
more strict calculation should consider the integral of $E$.
However, considering the uncertainty due to no clear plasma edge
like solid, the above approximation is enough for our estimation.
Using a dimensionless streamer density $\xi_t \equiv
\frac{4}{3}R_t^{3}n_0$ to the transverse direction, we have:
\begin{equation}\label{eq:densityt}
\xi_t \approx
\frac{4\pi}{3}\Big[\frac{4D_e}{\mu_eE_0\alpha(E_0)}\ln\frac{9\pi\varepsilon_0E_0}{2e\alpha^2(E_0)}\Big]^{3/2}{n_0}.
\end{equation}
The criterion for streamer coupling formation can be re-expressed
as:
\begin{equation}\label{eq:xicriterion}
\xi_t \geq 1,
\end{equation}
and thus we obtain the expression for primary electron density,
\begin{equation}\label{eq:ncriterion}
n_0 \gtrsim n_{\rm min} \equiv
\frac{3}{4\pi}\Big[\frac{4k_BT_e}{eE_0\alpha(E_0)}\ln\frac{9\pi\varepsilon_0E_0}{2e\alpha^2(E_0)}\Big]^{-3/2},
\end{equation}
where Einstein relation of $D_e/\mu_e = k_BT_e/e$ is used, $k_B$ is
the Boltzmann constant, and $T_e$ is the electron temperature.
$n_{\rm min}$ is defined as the minimum required preionization
level. Since $\alpha = \alpha(E_0, {\rm P})$, $n_{\rm min}$ is a
function of the external electric field $E_0$, the gas pressure
${\rm P}$ and the electron temperature $T_e$.

For the propagation of streamer coupling, the electric field
strength at the front of streamer head $E_M$ is estimated as:
\begin{equation}\label{eq:E}
E_M \approx \left\{\begin{array}{ll}
\Big(1 + \frac{9\pi}{4\alpha^3}n_0\Big)E_0&{\rm if}~\xi_p > 1,\\
1.5E_0&{\rm if}~\xi_p \leq 1,
\end{array} \right.
\end{equation}
where $\xi_p\equiv\frac{4\pi}{3}R_p^3n_0$. The strength of electric
field $E_M$ results from the total effect of streamer heads when
$\xi_p > 1$, and this effect disappears when $\xi_p \leq 1$.

\begin{figure}
\includegraphics[angle=0, width=0.4\textwidth]{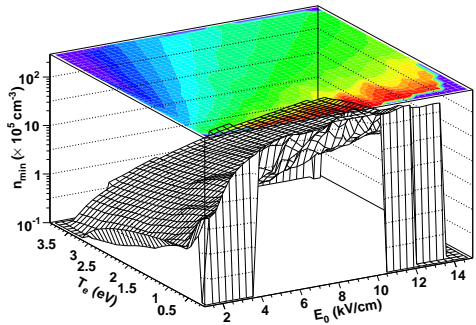}
\caption{\label{fig:density} Distribution of minimum required seed
electron density $n_{\rm min}$ for atmospheric pressure helium
plasma with electron temperature $T_e$ and electric field $E_0$. }
\end{figure}

Applying the above theoretical results to the atmospheric pressure
helium plasma, we find that the minimum required seed electron
density is relatively low for the discharge of a high electron
temperature and a low electric field (see Fig.~\ref{fig:density}).
Choosing data $\alpha \approx 5.3 \times 10^3 {\rm m^{-1}}$ from the
estimation of experimental value \cite{Data08}, and a typical
experimental condition of $T_e =2~{\rm eV}, E_0 = 4~{\rm kV/cm}$, we
can obtain:
\begin{equation}\label{eq:Hedensity}
n_{\rm min}^{\rm He} \sim 1.1 \times 10^{5}~{\rm cm^{-3}}\nonumber
\end{equation}
The above result is about one order of magnitude higher than the
predicted result from Palmer's model \cite{coupling74}. While the
experimental minimum required density for homogeneous discharge is
the order of $10^{5}~{\rm cm^{-3}}$ \cite{coupling80,coupling76},
which is in favor our predicted result.

Based on the above calculation, the predicted result for streamer
head radius in propagating direction is $R_p^{\rm He} \sim 0.03~{\rm
cm}$, and in transverse direction is $R_t^{\rm He} \sim 0.01~{\rm
cm}$ which is the radius of a single streamer. The two values
suggest that the distribution of space charge in streamer head likes
a "goose egg", as shown in Fig.~\ref{fig:streamer}(c).

\begin{figure}
\includegraphics[angle=0, width=0.5\textwidth]{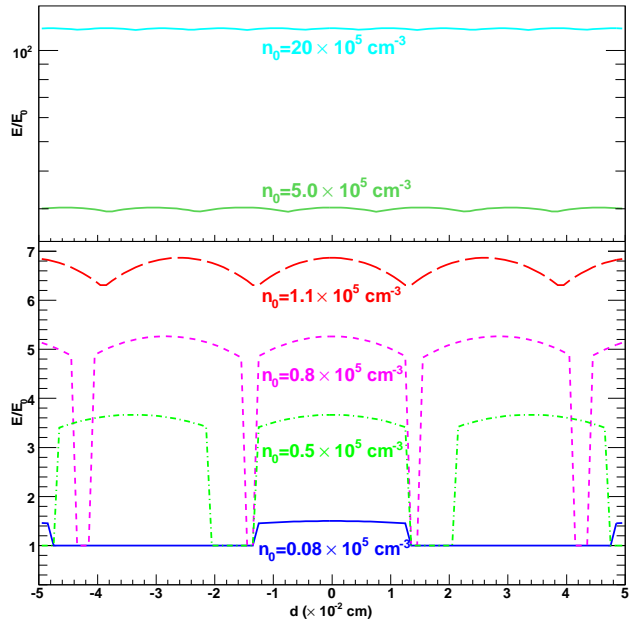}
\caption{\label{fig:field} Distribution of total electric field $E$
projected in streamer direction at the front of streamers for
different primary seed electron density. We choose electron
temperature of 2 eV and helium plasma of atmospheric pressure.}
\end{figure}

The transverse distribution of the total field strength $E$ in the
propagating direction at the front of streamer heads is shown in
Fig.~\ref{fig:field}. Streamers do not overlap with each other when
the primary seed electron density $n_0$ is lower than the minimum
required preionization level $n_{\rm min}$, such as $n_0 = 0.08$,
0.5 and $0.8 \times 10^5~{\rm cm^{-3}}$. The electric field is
continuous for the streamer coupling when $n_0 \geq n_{\rm min}$,
and its relative smoothness increases with $n_0$, such as $n_0 =
n_{\rm min},~5n_{\rm min}~{\rm and}~20n_{\rm min}$. This indicates
that the distribution of ionization and radiative processes are
almost homogeneous for the streamer coupling, and the emission
homogeneity is improved by increasing the preionization level $n_0$.
This prediction is qualitatively consistent with the experimental
results \cite{glow09,coupling80}, which suggest that the homogeneous
discharge can be only obtained with high preionization level.

\begin{figure}
\includegraphics[angle=0, width=0.4\textwidth]{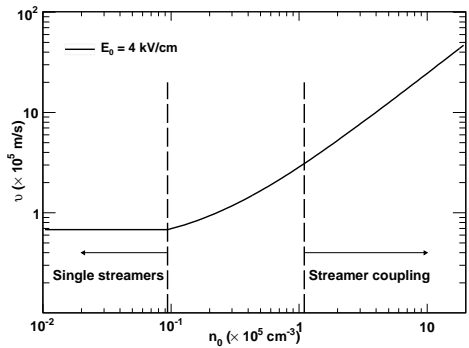}
\caption{\label{fig:speed} Distribution of the moving speed of
emission pattern with primary electron density at $T_e = 2$ eV and
$E_0 = 4~{\rm kV/cm}$ for atmospheric pressure plasma. The left and
right regions separated by vertical dashed lines denote discharges
of single streamers and streamer coupling, respectively.}
\end{figure}

The development of the streamer coupling is led by the drift of
electrons at the front of the streamer coupling head, since
streamers propagate along the direction of the strongest electric
field \cite{Raizer98}. Therefore we can obtain the moving speed of
the streamer coupling head:
\begin{equation}\label{eq:bulletspeed}
{\upsilon} \approx \left\{\begin{array}{ll}
\mu_eE_0\Big(1 + \frac{9\pi}{4\alpha^3}n_0\Big)E_0&{\rm if}~\xi_p > 1,\\
1.5\mu_eE_0&{\rm if}~\xi_p \leq 1.
\end{array} \right.
\end{equation}
Since the streamer coupling head is the most intensive ionization
region, the propagation of the streamer coupling head represents the
moving of discharge pattern in actual experiment. Using the data of
electron mobility $\mu_e = 1.1 \times 10^3~{\rm
cm^{2}V^{-1}{s^{-1}}}$ in the atmospheric pressure helium plasma
\cite{Raizer91}, we obtain the moving speed of the discharge pattern
which is shown in Fig.~\ref{fig:speed}. Different discharge regions
are separated by vertical dashed lines according to preionization
level. The left and right regions are single streamers, where
streamers are almost independent, and the streamer coupling, where
streamers overlap with each other, respectively. In the middle
region, although the discharge is also separated streamers, the
effect of other ones in transverse direction can not be ignored. It
shows that the moving speed of the discharge pattern increases
linearly with $n_0$ and $E_0$ in the streamer coupling region. The
scale of the moving speed is consistent with the experimental
results of "plasma bullet" \cite{bullet05,bullet10,La09,Lu06}.
Furthermore, according to Eq. (\ref{eq:bulletspeed}), the ionization
rate $\alpha$ and primary seed electron density $n_0$ is
axisymmetrically distributed due to the axisymmetrically distributed
of gas mixing and dielectric surface. The ionization rate $\alpha$
reaches its maximum value when the content ratio of nitrogen in the
helium plasma is at the level of $10^{-3}$ \cite{David10,ratio08}.
Therefore, the ring-shaped pattern of plasma bullet can also be
explained by Figs.~\ref{fig:field} and~\ref{fig:speed}.

Although the predicted results are consistent with done experiments,
an identifying experiment needs to be done to check the predictions
quantitatively. For the atmospheric pressure helium discharge, by
using the transversely excited atmospheric pressure ${\rm CO_2}$
laser system \cite{coupling76}, we can control the preionization
level $n_0$ in the setup of capacitively coupled plasma to identify
Figs.~\ref{fig:field} and \ref{fig:speed}. Fig.~\ref{fig:field}
suggests that the contrast of streamer emissions and background
increases with $n_0$ when $n_0 < n_{\rm min}$, and the homogeneity
and intensity of emission pattern increases with $n_0$ when $n_0 >
n_{\rm min}$. Fig.~\ref{fig:speed} suggests the moving speed of
discharge pattern under certain electric field and preionition
level, and that it increases with $n_0$.

To conclude, we have derived analytically the moving speed of
emission pattern and the minimum required preionization level for
anode-directed streamer coupling, and supposed that the streamer
coupling is required to generate a homogeneous emission pattern in
high pressure discharges. Both values depend on the electric field
strength and the gas pressure and the electron temperature. Based on
these predictions, we investigate the emission homogeneity and its
moving speed in atmospheric pressure helium plasma. Our predictions
are consistent with experimental results. The model of the streamer
coupling is very useful for understanding the dynamical process of
high pressure plasma.

We are very grateful to the help from Profs. Michael A. Lieberman,
Xinpei Lu and Michael G. Kong. Particular thanks are due to Drs.
Yukinori Sakiyama and Jiang-Tao Li for useful discussions.

\makeatletter
\def\bib@device#1#2{}
\makeatother


\end{document}